  \documentclass[sigconf,letterpaper,natbib=false]{acmart}
\usepackage{makecell}
\usepackage[linesnumbered,boxed,nofillcomment,scleft]{algorithm2e}
\usepackage{subcaption}
\usepackage{xspace}
\usepackage{multirow}
\usepackage{pifont}
\usepackage{colortbl}

\usepackage{tikz}
\usetikzlibrary{arrows,positioning}

\tikzstyle{dnsname}=[right,rectangle,fill,minimum size=3em,scale=.9,solid,fill=green!10,minimum height=1em]
\tikzstyle{dnsnameB}=[right,rectangle,fill,minimum size=1em,scale=.9,solid,fill=green!40,minimum height=1em]
\tikzstyle{linkedname}=[right,rectangle,fill,minimum size=3em,scale=.9,solid,fill=blue!10,minimum height=1em]
\tikzstyle{linkednameB}=[right,rectangle,fill,minimum size=3em,scale=.9,solid,fill=blue!40,minimum height=1em]
\tikzstyle{zombiename}=[right,rectangle,draw,dashed,minimum size=3em,scale=.9,solid,fill=red!20,minimum height=1em]
\tikzstyle{zombienameB}=[right,rectangle,minimum size=3em,scale=.9,fill=red!60,minimum height=1em]

\def\Snospace~{\S{}}
\hypersetup{
    colorlinks,%
    citecolor=[rgb]{0.7,0,0},
    filecolor=[rgb]{0.2,0.2,0.2},
    linkcolor=[rgb]{0,0,0.7},
    urlcolor=[rgb]{0.7,0,0},
}

\makeatletter
  \newcommand\EatSpacesHack{\@bsphack\@esphack}
\makeatother

  \renewcommand\comment[1]{\EatSpacesHack}
  \newcommand\PostSubmission[1]{\EatSpacesHack}
  \newcommand\STV[2][]{\EatSpacesHack}
  \newcommand\reviewfix[1]{\EatSpacesHack}
\newcolumntype{P}[1]{>{\raggedright\arraybackslash}p{#1}}

\makeatletter
  \renewcommand\section{\def\@toclevel{1}%
    \@startsection{section}{1}{\z@}%
    {-.4\baselineskip \@plus -2\p@ \@minus -.2\p@}%
    {.2\baselineskip}%
    {\ACM@NRadjust\@secfont}}
  \let\ACM@origsection\section

  \renewcommand\subsection{\def\@toclevel{2}%
    \@startsection{subsection}{2}{\z@}%
    {-.25\baselineskip \@plus -2\p@ \@minus -.2\p@}%
    {.1\baselineskip}%
    {\ACM@NRadjust\@subsecfont}}
  \let\ACM@origsubsection\subsection

  \renewcommand\subsubsection{\def\@toclevel{3}%
    \@startsection{subsubsection}{3}{\z@}%
    {-.25\baselineskip \@plus -2\p@ \@minus -.2\p@}%
    {-0.01\p@}%
    {\ACM@NRadjust{\@subsubsecfont}}}
  \let\ACM@origsubsubsection\subsubsection
\makeatother
  \copyrightyear{2026}
  \setcopyright{rightsretained}
  \acmDOI{}
  \acmConference{Draft (Under Submission)}{May~2026}{Los Angeles, CA, USA}
  \acmISBN{}
\RequirePackage[datamodel=acmdatamodel,style=acmnumeric]{biblatex}

\begin{document}

\title[ %
  Zombies in Alternate Realities: The Afterlife of Domain Names in DNS Integrations
]{ %
  Zombies in Alternate Realities:\\
  The Afterlife of Domain Names in DNS Integrations
}

 % \ifisanon
 % \else
  \author{Sulyab Thottungal Valapu}
  \orcid{0000-0001-6332-1344}
  \affiliation{
      \institution{USC/Information Sciences Institute}
      \city{Los Angeles}
      \country{United States}
  }
  \email{sulyab.tv@usc.edu}
  
  \author{John Heidemann}
  \orcid{0000-0002-1225-7562}
  \affiliation{
      \institution{USC/Information Sciences Institute}
      \city{Los Angeles}
      \country{United States}
  }
  \email{johnh@isi.edu}

  \author{Mattijs Jonker}
  \affiliation{
      \institution{University of Twente}
      \city{Enschede}
      \country{The Netherlands}
  }

  \author{Raffaele Sommese}
  \affiliation{
      \institution{University of Twente}
      \city{Enschede}
      \country{The Netherlands}
  }
  %
 % \fi

%
%
%
%
%
%
 % \ifisanon
 % \else
\renewcommand{\shortauthors}{Sulyab Thottungal Valapu et al.}
 % \fi

%
\begin{abstract}
DNS integrations
  leverage the discovery, trust, %
  and uniqueness
  of the global Domain Name System
  with a \emph{linkage} to another naming ecosystem,
  so the DNS name can help identify resources such as 
  a cryptocurrency wallet or software component.
While DNS ownership is verified at linkage creation,
  many ecosystems do not track subsequent DNS changes.
The result is \emph{zombie linkages}, 
  where the DNS ownership has expired or changed,
  but the mapping to the linked %
  resource persists.
We define a threat model for DNS integrations, identifying
  five %
  classes of attacks that leverage or exploit zombie linkages.
We measure zombie occurrence 
  across three %
  DNS integrations---Web PKI; %
  ENS, a blockchain naming system; %
  and Maven Central, a Java software repository.
We show that zombies exist in every ecosystem,
  but at very different fractions---zombies make up roughly
  3\% of TLS certificates for new domains,
  24\% of ENS on-chain imports,
  and 15\% of Maven Central namespaces.
We evaluate how integration design choices affect outcomes,
  with validate-once integrations (ENS on-chain, Maven Central)
  accumulating long-lasting zombies,
  linkages with expiration (Web PKI) limiting damage,
  while integrations that validate on every use (ENS gasless) are zombie-free by design.
We look for specific attacks, finding
  attacks actively available for exploitation
  in both Web PKI and Maven Central.
Finally, we recommend steps to reduce zombie occurrence.
\end{abstract}


%
%
%

%
%
%
 % \ifistechreport
 % \else
 % \fi

%
%
%

%
%
%
\maketitle

\section{Introduction}
	\label{sec:intro}

DNS, the Internet's Domain Name System~\cite{Mockapetris83a},
  provides a universal namespace for the Internet.
The importance of the DNS has given rise to a rich ecosystem
  of registries that coordinate use of top-level domains (TLDs),
  and registrars that sell names to the public.
Together, they allocate domain names to individual owners,
  giving each each owner authority over resources under that domain.

\emph{DNS integration}
  is when a new, independent naming ecosystem
  links itself to DNS names~\cite{sheth_integration_2025}.
They do so to leverage the fact that domain names in DNS
  are globally unique, human-friendly,
  and already widely recognized.
The Web Public Key Infrastructure (Web PKI),
  one of the most widely deployed integrations,
  binds TLS certificates to DNS names
  to secure web communications~\cite{Dierks08a}.
Other integrations include
  the Ethereum Name Service (ENS),
  which lets users associate DNS names
  with cryptocurrency wallets~\cite{mcdee_ensip-6_2018};
  Maven Central, which derives
  Java package namespaces from DNS names,
and
  Bluesky, which supports using DNS names
  as social media handles~\cite{graber_domain_2023}.

For all these benefits,
  integrating with DNS comes with a cost:
  \emph{DNS names change hands, and linked ecosystems must keep up}.
Domains expire, get transferred,
  or are re-registered by different parties,
  and no mechanism exists for ecosystems
  to subscribe to such changes in DNS state.
While some periodically re-validate the DNS linkage, not all do.

\emph{When integrations become desynchronized, 
  linkages become zombies, creating new risks}
  as the linkages outlive the DNS ownership epoch that created them
  (\autoref{sec:problem_statement/zombies}).
Consequences of zombie linkages range from user confusion
  to exploitable security vulnerabilities (\autoref{sec:problem_statement/attacks}).
If an adversary squats on a linked name backed by a DNS name that is later re-registered,
  they may trick users into sending cryptocurrency
  to the wrong party.
If an adversary acquires an established DNS name that expired,
  they may take over established linked resources such as software packages
  and mislead users into installing malware.

We provide the first systematic examination of
  the security threats that emerge
  when DNS integrations fail to track
  DNS ownership changes.
Prior work has studied
  synchronization challenges 
  in individual integrations~\cite{ma_stale_2023,kaizer_synchronization_2025},
  (see also \autoref{sec:related}).
Different ecosystems take different approaches:
  proactively limiting integration lifetime with built-in expiration,
  reactively removing stale records on discovery,
  or neither.
Our goal is to understand what defenses are used and how effective they are.
As the number of DNS integrations grows,
  exploring these questions grows more urgent.

This paper makes four contributions:
First, we define the problem and develop a threat model 
  when DNS integrations become desynchronized
  (\autoref{sec:problem_statement}).
Second, we measure the prevalence and persistence
  of exploitable zombie linkages
  across three diverse integrations (\autoref{sec:methodology}):
  Web PKI (web security),
  ENS (cryptocurrency),
  and Maven Central (software package repository),
  and report the results (\autoref{sec:char-zombies}).
Third, we examine whether zombies
  translate into exploitable attack surface
  in practice (\autoref{sec:attacks}).
Finally, drawing on these findings,
  we recommend steps to reduce zombie occurrence (\autoref{sec:discussion}).

We measure zombies across every ecosystem we study,
  finding sharply different rates, determined by ecosystem design (\autoref{sec:char-zombies}).
validate-once integrations accumulate the highest fraction of zombies:
  24\% of ENS On-chain imports and 15\% of Maven Central namespaces
  are zombies, with none ever cleared in ENS On-chain
  and all permanent in Maven Central.
Web PKI's bounded certificate lifetimes
  keep the zombie fraction at 2.7\% for newly created domains,
  and the rarity of revocation (4.3\% zombie certificates revoked)
  suggests lower certificate lifetimes as the most effective defense.
We find that zombie formation is driven by routine DNS name expiration
  in ENS and Maven Central,
  whereas in Web PKI we see extremely short-lived linkages consistent with abuse.

We look for attacks involving zombies that are actively available for exploitation (\autoref{sec:attacks}),
  and find over 192k Web PKI zombie certificates
  still served months after DNS name death,
  7.3k of which continue to be served past DNS re-registration,
  and 214 Maven Central zombie namespaces
  with publishing activity after DNS re-registration.
Our findings show that zombie linkages
  are actively available for domain impersonation
  and software supply chain attacks.

\textbf{Ethical Considerations and Data Availability:}
This paper used a range of non-personal data
  (DNS zone files, RDAP queries, TLS certificates,
  Maven Central metadata,
  and Ethereum records),
  most of it available publicly.
Our work uses no personally identifiable information that is not already public.
We assess security threats, but as general types of attacks,
  so our work does not require coordinated disclosure.
We discuss ethics in detail in \autoref{sec:appendix/ethics}.

This paper uses data from third parties and newly collected data.
Data from third-parties
  (dns.coffee~\cite{dnscoffee} and OpenINTEL~\cite{van_rijswijk-deij_high-performance_2016,sommese_darkdns_2024})
  is used with their permission and must be obtained from them.
  %
%
%
 % \ifisanon
 % \else
Data we have collected and pointers to third-party data
  is listed on our website~\cite{ThottungalValapu26b}.
 % \fi

%
\section{Related Work}
  \label{sec:related}

To our knowledge, only one group has previously explored the general
  challenge of DNS integrations,
  but several groups have explored specific integrations in 
  the specific ecosystems we study
  (domain re-registration, Web PKI, blockchain naming systems, software package management).

To our knowledge, only Verisign has studied 
  DNS integrations as class of security risks~\cite{sheth_call_2023,kaizer_poster_2024,kaizer_synchronization_2025}.
They 
  categorize DNS-to-ecosystem name imports as
  synchronized, not synchronized, or ambiguous
  and study ENS On-chain, Bluesky, and GitHub Organizations.
Our work was inspired by their study,
  but goes further to describe the threat model,
  characterizing zombie lifetimes and formation patterns,
  and begin the search for evidence of exploitation.

Several studies have examined
  the security risks of expired and re-registered domains.
Problems here include
  residual trust and its abuse in re-registered DNS domains~\cite{lever_domain-z_2016},
  and residual traffic in re-registered domains~\cite{so_domains_2022}.
Dropcatching is acquiring domains when they expire,
  either to auction, or to provide advertising or malware~\cite{miramirkhani_panning_2018,lauinger_game_2017}.
These prior works focus on DNS-specific implications of re-registration,
  we extend this work to consider \emph{external ecosystems}
  that integrate with DNS.

In Web PKI,
  Ma et al. study stale TLS certificates
  over a ten-year period,
  finding that infrastructure changes leave third parties
  with TLS private keys for domains they no longer control,
  suggesting  shorter certificate lifetimes~\cite{ma_stale_2023}.
Our results complement theirs:
  we focus on zombie certificates caused by DNS ownership changes
  and measure how long they continue to be served
  after DNS name death and re-registration,
  also finding revocation unreliable as a defense.

In blockchain naming,
  researchers have studied ENS adoption and identify squatting
  and malicious content~\cite{xia_challenges_2022},
  and study dropcatching of ENS-native \texttt{.eth} names
  to misdirect funds~\cite{muzammil_panning_2024}.
These studies address ENS's own namespace;
  we expand on both of these attacks 
  and consider DNS integration as a broader risk.
Other work has examined TLD collisions between
  blockchain naming systems and DNS~\cite{randall_challenges_2022,ito_investigations_2024},
  work orthogonal to ours.

In software supply chains,
  security researches identified DNS reallocation 
  as a supply-chain risk for Maven with ``MavenGate''~\cite{oversecured_introducing_2024}.
We expand on and quantify this attack in \autoref{sec:attacks/maven}.
While they identify vulnerable namespaces
  from public domain availability,
  we use DNS registration history
  to identify namespaces that have \emph{already} undergone
  ownership changes and exhibit post-re-registration
  publishing activity.
Maven is just one software packaging system;
  Gu et al. study six ecosystems and
  identify twelve attack vectors~\cite{gu_investigating_2023};
  but they do not consider DNS ownership changes as an attack class,
  our focus.

Overall, our work is first work to develop
  a unified threat model for DNS integrations,
  characterize zombie linkages across multiple ecosystems,
  consider how ecosystem design choices produce different zombie behavior,
  and identify attack patterns through analyzing the zombie lifecycle.
\section{Problem Statement}
  \label{sec:problem_statement}

\begin{table*}
\centering
\resizebox{0.85\textwidth}{!}{
  \begin{tabular}{P{1in}|P{1.4in}|P{2in}|P{1.4in}}
  \hline
  \textbf{Term} & \textbf{Web PKI} & \textbf{ENS} & \textbf{Maven Central} \\
  \hline
  DNS name & \multicolumn{3}{c}{\texttt{example.com}} \\
  \hline
  Linked name & Public key & Wallet address & \texttt{com.example} namespace \\
  \hline
  Linked resource & Authenticated website & Ethereum wallet \newline (balance, txn history) & Packages published under the namespace \\
  \hline
  Resource \newline controlled by & TLS private key holder & Wallet private key holder & Account holder that completes the linkage process \\
  \hline
  Linkage process & CA issues certificate after domain control validation & \emph{On-chain:} User submits Ethereum transaction with DNSSEC proof to ENS smart contract \newline \emph{Gasless:} User turns on DNSSEC; sets up DNS TXT record & User proves DNS ownership via DNS TXT record \\
  \hline
  Linkage entry & Certificate & \emph{On-chain:} Blockchain transaction \newline \emph{Gasless:} DNS TXT record & Database record \\
  \hline
  Linkage entry \newline lives in & Web server, CT logs & \emph{On-chain:} Ethereum blockchain \newline \emph{Gasless:} DNS & Maven Central's internal database \\
  \hline
  \end{tabular}
}
\caption{
  Terms applied to each DNS integration we study.
}
\label{tab:terms}
\end{table*}

Our goal is to understand how linkages
  created in non-DNS ecosystems
  on the basis of DNS ownership
  can outlive that ownership,
  and how that can lead to abuse.

\subsection{How Ecosystems Integrate DNS}
  \label{sec:problem_statement/integrations}

We use \emph{ecosystem} to refer to
  any non-DNS system that maintains its own namespace.
\autoref{tab:terms} shows the three ecosystems we study:
  Web PKI~\cite{Dierks08a},
  ENS (Ethereum Name Service)~\cite{mcdee_ensip-6_2018},
  and Maven Central~\cite{maven_central_about_nodate}.

A \emph{DNS integration} is a mechanism by which
  an ecosystem uses a DNS name as an identifier
  within its own namespace~\cite{sheth_integration_2025}.
DNS names are attractive for this purpose
  because they are human-friendly,
  unlike native identifiers in some ecosystems,
  and because they ground new namespaces
  in the most widely deployed
  namespace on the Internet.

Each integration binds a \emph{DNS name}
  to an identifier in the ecosystem,
  which we call the \emph{linked name}.
Linked names often support additional information or capability,
  which we call the \emph{linked resource}.
For example, in ENS, a DNS name is linked
  to an Ethereum wallet address (the linked name),
  and the linked resource is the contents of wallet itself.

The linkage from DNS to the ecosystem is established
  through a \emph{linkage process}
  in which the ecosystem verifies
  that the requesting user controls the DNS name.
The exact mechanism for this linkage process differs by ecosystem,
  but it typically involves a challenge to
  demonstrate control of the DNS name.
For example, the ecosystem may require that
  the DNS name owner demonstrate control
  by creating a specific DNS TXT record,
  and then it verifies the presence of this record
  to establish the linkage.

Completing a linkage successfully
  creates a \emph{linkage entry}:
  a persistent artifact in the ecosystem
  that maps a DNS name to a linked name.
In some ecosystems, such as Web PKI,
  linkage entries automatically expire after
  some included time-to-live.
In others, such as ENS On-chain,
  linkage entries remain valid indefinitely until explicitly removed.

\autoref{tab:terms} relates each term to the three integrations we study.

\subsection{Background About Studied Ecosystems}

We study three ecosystems here,
  and identify several other ecosystems for future study.

\textbf{Web PKI.}
Web PKI is the cryptographic infrastructure used to secure the web
  and other TLS-based communication~\cite{Dierks08a}.
Its linkage entries are X.509 certificates
  binding DNS names to public keys (linked name)
  via the certificate's Common Name
  and Subject Alternative Name fields.
In the ecosystem, 
  certificates are signed by other certificates in a hierarchy 
  rooted in
  trusted Certificate Authorities provided with
  the operating system or web browser.
We focus on domain-validated certificates,
  the most common, making up
  95\% of CT logs~\cite{cloudflare_certificate_nodate}.

\textbf{ENS.}
ENS uses Ethereum
  to map human-readable names to blockchain addresses~\cite{mcdee_ensip-6_2018}.
While ENS has its own \texttt{.eth} names,
  we focus on its linking DNS names
  to Ethereum wallet addresses (linked name) and wallets (the linked resource).

ENS supports two linkage methods.
Since 2018,
  \emph{On-chain} linkages
  write an entry to the Ethereum blockchain,
  including a DNSSEC-proof with the \texttt{DNSRegistrar} smart contract.
Since 2024, \emph{Gasless} linkages no longer modify the blockchain,
  but instead validate a
  DNSSEC-signed TXT record
  on each use~\cite{gregskrileth_gasless_2024}.
We describe both as \emph{ENS},
  adding \emph{On-chain} and \emph{Gasless} when needed.

\textbf{Maven Central.}
Maven Central is a public repository
  of Java software components~\cite{maven_central_about_nodate},
  used automatically by build tools such as Apache Maven~\cite{apache_software_foundation_introduction_nodate}.
Anyone can publish new components (the linked resource)
  after creating an account on the web portal
  and registering a namespace (the linked name)
  verified through a DNS TXT record.

Maven Central's components are immutable
  once published~\cite{sonatype_immutability_nodate},
  although they occasionally remove malware~\cite{sonatype_security_research_team_sonatype_2021}.
DNS integration and package immutability results in attacks
  we study in \autoref{sec:attacks/maven}.

\subsection{How Linkages Become Zombies}
  \label{sec:problem_statement/zombies}

\begin{figure}
%
% xxx: this was my first picture 
% so it uses absolute coordinates.
% look at bulk_linked_name_creation.tex for better code
% with relative positioning
%
\begin{tikzpicture}[node distance=2.5cm,auto,>=latex',scale=0.85, transform 
shape]
  \node[right] at (0,0) {\textbf{DNS Name}};
  \node [dnsname,text width=3cm] (dnsname) at (3,0) { \texttt{example.com} };
  \draw [->,thick] (dnsname.south west) -- (dnsname.north west)+(0,0.2) node[right] {\small \emph{birth}}; % birth
  \draw [->,thick] (dnsname.north east)+(0,0.2) -- (dnsname.south east) node[very near end, right] {\small \emph{withdrawal}}; % death
  \draw [->,dashed] (4,-.2) to [bend right] (4,-2); %  node[midway] {\small \emph{linkage}}; % linkage
  \node[fill=none,right] at (3.8,-1) {\small \emph{linkage}};  % xxx: sigh, using bend right makes the midwaynode go away

  \node[right] at (0,-2) {\textbf{Linked Name}};
  \node [linkedname,text width=2cm] (linkedname) at (4,-2) { \texttt{ens:0x123456}\hspace*{4em} };
  \draw [->,thick] (linkedname.south west) -- (linkedname.north west)+(0,0.2); % birth
  \node [zombiename,text width=1.5cm] (zombielinkedname) at (6,-2) { \emph{(zombie)} };
  \draw [->,thick,red] (zombielinkedname.south west) --  (zombielinkedname.north west)+(0,0.2); % zombie birth
  \node[fill=none,right] at (5.9,-1.2) {\small \emph{(zombie}};
  \node[fill=none,right] at (5.9,-1.5) {\small \emph{birth)}};
  \draw [->,thick] (zombielinkedname.north east)+(0,0.2) -- (zombielinkedname.south east) node[at end, right] {\small \emph{death}}; % death

\end{tikzpicture}
\caption{Timing during name integration via linkage.}
\end{figure}
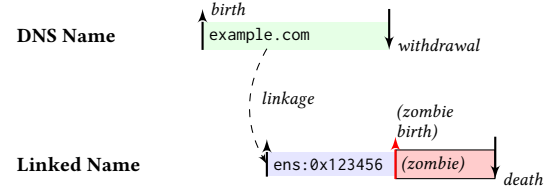

DNS name ownership is not permanent.
Domains expire, are transferred,
  or are re-registered by different parties.
We define a \emph{DNS ownership epoch}
  as a period of continuous control
  of a DNS name by a single user.

A linkage becomes a \emph{zombie}
  when the DNS ownership epoch
  during which it was created has ended,
  but its linkage entry remains valid in the ecosystem.
A zombie linkage continues to map a DNS name
  to the original linked name,
  an assertion that is no longer valid.

How long a zombie persists
  depends on two properties of the ecosystem's design:
  the issued lifetime of the linkage entry,
  and the ecosystem's revocation mechanism.

If a linkage entry is issued
  with a finite validity period,
  that period acts as an upper bound on zombie duration:
  a zombie TLS certificate cannot outlive its expiration date.
Ecosystems that issue linkage entries
  without lifetime constraints
  lack this natural bound.

Within this bound, zombie duration depends on
  whether the ecosystem revokes zombies,
  and how quickly.
Some ecosystems, such as ENS On-chain,
  have no revocation mechanism at all,
  leaving zombies to persist indefinitely.
Others do revoke, but the delay
  between DNS ownership change and revocation
  determines the window of exposure.

\subsection{Attacks on DNS Integrations}
  \label{sec:problem_statement/attacks}

Integration between DNS and other ecosystems raises the problem
  that they can fall out of synchronization, presenting new threats.
We look at five types of attack,
  with three focusing on the DNS side:
  Bulk Linked Name Creation,
  Linked Name Squatting,
  and Linked Name Takeover.
The two other attacks occur in the linked ecosystem:
 Linked Resource Squatting,
 and Linked Resource Takeover.

\emph{Squatting attacks} are when an adversary acquires
  a \emph{new} linked name or resource,
  anticipating that it will have value in the future.
We generalize this term from domain squatting~\cite{cloudflare_what_nodate},
  to naming integration.
\emph{Takeover attacks} are when an adversary
  takes control of an existing linked name or resource
  with the hope of gaining value from it.
This term follows from
  subdomain takeover attacks~\cite{mdn_contributors_subdomain_2025}.

\begin{figure}
\subfloat[Bulk Linked Name Creation]{
  \label{fig:bulk_linked_name_creation_picture}
\begin{tikzpicture}[node distance=2.5cm,auto,>=latex',scale=0.85, transform 
shape]
  \colorlet{dnscolor}{green!60!black}
  \colorlet{linkcolor}{blue!70!black}
  \colorlet{zombiecolor}{red!75!black}
  \node[right] at (-3.25,3) {\emph{all adversary owned}};
  \foreach \pos in {0,...,1} {
    \node at (0.1*\pos,2*\pos) {\tikz{

    \node[right] (dnsnametitle) at (0,0) {\textbf{DNS Name}};
    \node[below=0.5cm of dnsnametitle.south west,right]  {\textbf{Linked Name}};
  
    \node [dnsnameB,text width=0.05cm,minimum size=1.5em] (dnsname) at (3,0) {\hspace*{1.5em}\vphantom{Xy}};
    \draw [->,very thick,dnscolor] (dnsname.south west) -- ([yshift=0.2cm]dnsname.north west);
    \draw [->,very thick,dnscolor] (dnsname.north east) -- ([yshift=-0.2cm]dnsname.south east);

    \node[linkednameB,text width=0.05cm,below right=0.5cm and 0.2cm of dnsname.south west] (linkedname) {\vphantom{Xy}};
    \node [zombienameB,text width=2cm,below right=0.5cm and 0cm of dnsname.south east] (zombielinkedname) {\vphantom{Xy}example\pos.com};
    \draw [->,very thick,linkcolor] (linkedname.south west) -- ([yshift=0.2cm]linkedname.north west); %
    \draw [->,very thick,zombiecolor] (zombielinkedname.south west) -- ([yshift=0.2cm]zombielinkedname.north west); %
    \draw [->,very thick,zombiecolor] (zombielinkedname.north east) -- ([yshift=-0.2cm]zombielinkedname.south east); %
  
    \draw [->,dashed] ([xshift=0.2cm]dnsname.south west) to [bend right] (linkedname.south west); %

    \draw (dnsname.east)+(0.6,0) node[rotate=90] {\small \emph{(AGP)}};

    }};

  }
  \node at (0.3,-1.2) {
    \tikz{
      \foreach \i in {0,...,2} {
        \draw[fill=black] (0,\i*0.2cm) circle (0.05cm);
      };
    }
  };
\end{tikzpicture}
} \\
\subfloat[Linked Name or Resource Squatting]{
  \label{fig:linked_name_squatting_picture}
\begin{tikzpicture}[node distance=2.5cm,auto,>=latex',scale=0.85, transform 
shape]
    \node[right] (dnsnametitle) at (0,0) {\textbf{DNS Name}};
    \node[below=0.8cm of dnsnametitle.south west,right]  {\textbf{Linked Name}};
  
    \node [dnsnameB] (dnsname) at (3,0) { target.com };
    \node [above=0cm of dnsname.north west, anchor=south west] { \small \emph{adver. owned} };
    \draw [->,thick] (dnsname.south west) -- (dnsname.north west)+(0,0.2);
    \draw [->,thick] (dnsname.north east)+(0,0.2) -- (dnsname.south east);

    \node [dnsname,right=2cm of dnsname.east,text width=3cm] (dnsnamevictim) at (3,0) { target.com \emph{(re-alloc.)} };
    \draw [->,thick] (dnsnamevictim.south west) -- (dnsnamevictim.north west)+(0,0.2);
    \node [above=0cm of dnsnamevictim.north west,anchor=south west] { \small\emph{victim owned} };

    \node[linkednameB,text width=2cm,below right=0.8cm and 0.2cm of dnsname.south west] (linkedname) { \vphantom{Xy} };
    \node [zombienameB,text width=3.5cm,below right=0.8cm and 0cm of dnsname.south east] (zombielinkedname) { target linked name };
    \node [above=0cm of linkedname.north west,anchor=south west] { \small \emph{adversary owned} };
    \draw [->,thick] (linkedname.south west) -- (linkedname.north west)+(0,0.2); %
    \draw [->,thick,red] (zombielinkedname.south west) --  (zombielinkedname.north west)+(0,0.2); %

    \draw [->,dashed] (dnsname.south west)+(0.2,0) to [bend right] (linkedname.south west); %
\end{tikzpicture}
}  \\
\subfloat[Linked Name Takeover]{
  \label{fig:linked_name_takeover_picture}
\begin{tikzpicture}[node distance=2.5cm,auto,>=latex',scale=0.85, transform 
shape]
    \node[right] (dnsnametitle) at (0,0) {\textbf{DNS Name}};
    \node[below=2.2cm of dnsnametitle.south west,right]  {\textbf{Linked Name}};
    \node[below=1cm of dnsnametitle.south west,right]  {\textbf{Adversary's Linked Name}};
  
    \node [dnsname] (dnsname) at (3,0) { target.com };
    \node [above=0cm of dnsname.north west, anchor=south west] { \small \emph{victim} };
    \draw [->,thick] (dnsname.south west) -- (dnsname.north west)+(0,0.2);
    \draw [->,thick] (dnsname.north east)+(0,0.2) -- (dnsname.south east);

    \node [dnsnameB,right=2cm of dnsname.east,text width=3cm] (dnsnameadv) at (3,0) { target.com \emph{(re-alloc.)} };
    \draw [->,thick] (dnsnameadv.south west) -- (dnsnamevictim.north west)+(0,0.2);
    \node [above=0cm of dnsnameadv.north west,anchor=south west] { \small\emph{adversary owned} };

    \node[linkedname,text width=2cm,below right=2cm and 0.2cm of dnsname.south west] (linkedname) { \vphantom{Xy} };
    \node [zombiename,text width=3.5cm,below right=2cm and 0cm of dnsname.south east] (zombielinkedname) { target linked name };
    \node [above=0cm of linkedname.north west,anchor=south west] { \small \emph{victim owned} };
    \draw [->,thick] (linkedname.south west) -- (linkedname.north west)+(0,0.2); %
    \draw [->,thick,red] (zombielinkedname.south west) --  (zombielinkedname.north west)+(0,0.2); %

    \draw [->,dashed] (dnsname.south west)+(0.2,0) to [bend right] (linkedname.south west); %

    \node[linkednameB,text width=3cm,below right=0.8cm and 0.2cm of dnsnameadv.south west] (linkednameadv) { \small adversary's lnkd.~nm. };
    \node [above=0cm of linkednameadv.north west,anchor=south west] { \small \emph{adversary owned} };
    \draw [->,thick] (linkednameadv.south west) -- (linkednameadv.north west)+(0,0.2); %

    \draw [->,dashed] (dnsnameadv.south west)+(0.2,0) to [bend right] (linkednameadv.south west); %
\end{tikzpicture}
}  \\
\subfloat[Linked Resource Takeover]{
  \label{fig:linked_resource_takeover_picture}
\begin{tikzpicture}[node distance=2.5cm,auto,>=latex',scale=0.85, transform 
shape]
    \node[right] (dnsnametitle) at (0,0) {\textbf{DNS Name}};
    \node[below=0.8cm of dnsnametitle.south west,right]  {\textbf{Linked Name}};
  
    \node [dnsname] (dnsname) at (3,0) { target.com };
    \node [above=0cm of dnsname.north west, anchor=south west] { \small \emph{victim} };
    \draw [->,thick] (dnsname.south west) -- (dnsname.north west)+(0,0.2);
    \draw [->,thick] (dnsname.north east)+(0,0.2) -- (dnsname.south east);

    \node [dnsnameB,right=2cm of dnsname.east,text width=3cm] (dnsnameadv) at (3,0) { target.com \emph{(re-alloc.)} };
    \draw [->,thick] (dnsnameadv.south west) -- (dnsnamevictim.north west)+(0,0.2);
    \node [above=0cm of dnsnameadv.north west,anchor=south west] { \small\emph{adversary controlled} };

    \node[linkedname,text width=2cm,below right=0.8cm and 0.2cm of dnsname.south west] (linkedname) { target lnk. };
    \node [zombiename,text width=3.5cm,below right=0.8cm and 0cm of dnsname.south east] (zombielinkedname) { \vphantom{Xy}nm. };
    \node [above=0cm of linkedname.north west,anchor=south west] { \small \emph{victim cntled.} };
    \draw [->,thick] (linkedname.south west) -- (linkedname.north west)+(0,0.2); %
    \draw [->,thick,red] (zombielinkedname.south west) --  (zombielinkedname.north west)+(0,0.2); %

    \draw [->,dashed] (dnsname.south west)+(0.2,0) to [bend right] (linkedname.south west); %

    \node[linkednameB,text width=3cm,below right=0.8cm and 0.2cm of dnsnameadv.south west] (linkednameadv) { \small adversary's lnkd.~nm. };
    \node [above=0cm of linkednameadv.north west,anchor=south west] { \small \emph{adversary controlled} };
    \draw [->,thick] (linkednameadv.south west) -- (linkednameadv.north west)+(0,0.2); %

    \draw [->,dashed] (dnsnameadv.south west)+(0.2,0) to [bend right] (linkednameadv.south west); %
\end{tikzpicture}
} 
\caption{Different attacks on naming integration.}
\end{figure}
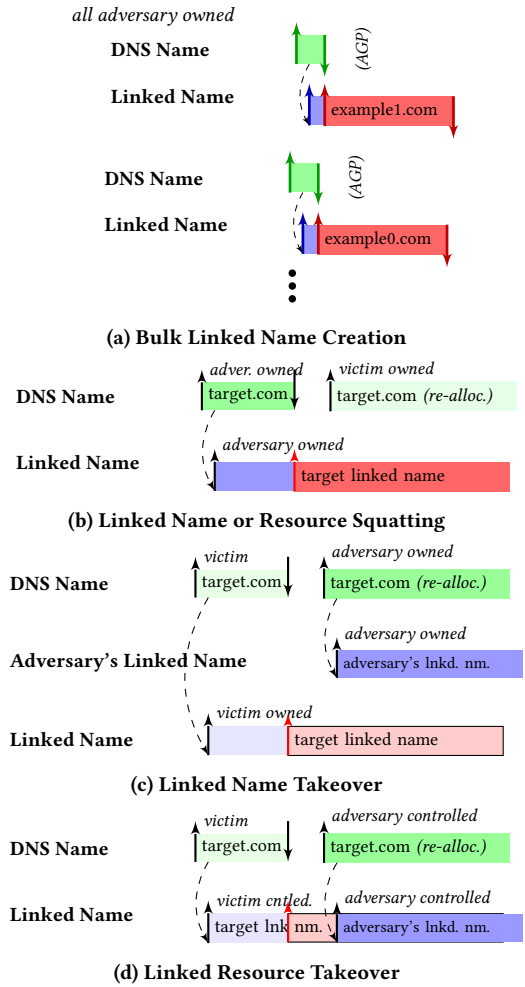

\subsubsection{Bulk Linked Name Creation}\:
  is the first type of squatting attack
  (\autoref{fig:bulk_linked_name_creation_picture}).
In this case the adversary acquires many DNS names at low cost.
The adversary then immediately links them into another ecosystem,
  creating the linked name and resource,
  and then quickly deletes the DNS names.
DNS names can be created at very low cost
  using \emph{domain tasting}~\cite{dynadot_what_2025},
  where names discarded during
  Add Grace Period (AGP)~\cite{icann_agp_2008}
  can be tried for as little as
  \$0.10~USD per DNS name~\cite{dynadot_grace_nodate},
  plus any ecosystem-specific fees.
Although ICANN limits the number of free AGP deletions
  per registrar per month~\cite{icann_end_2009},
  an adversary can distribute the attack
  across multiple registrars or over several months.

Once the adversary amasses linked names in bulk for cheap,
  they can be used in spamming campaigns,
  laundering data through a chain of names, 
  or for other throw-away purposes.

The exact victim of this attack depends on how the linked names are used.
For example, an adversary who retains
  unrevoked TLS certificates
  for DNS names that have since been suspended for abuse
  and who can direct victims to a server
  (such as through DNS cache poisoning~\cite{jiang_ghost_2012,li_ghost_2023})
  can serve content that browsers accept as authenticated.
The adversary can then serve phishing or malware pages,
  or run trojan attacks on these victims.

\subsubsection{Linked Name Squatting}\:
  is when the adversary creates a DNS name
  they expect someone else will later desire,
  so the adversary can hold the linked name.
If the ecosystem does not revoke the linkage entry
  when DNS name ownership ends,
  the result is a zombie linkage.
There are two subcases of this attack 
  depending on whether the DNS name is re-registered or not.

If the DNS name goes unclaimed (\autoref{fig:bulk_linked_name_creation_picture}), as is typical in Bulk Linked Name Creation,
  the adversary exploits residual trust
  in the linked name.

In the second case (\autoref{fig:linked_name_squatting_picture}),
  the DNS name is re-registered by another party,
  but this new DNS name owner coexists with
  the zombie linked name held by the adversary.
The adversary can then impersonate
  the new owner in the ecosystem
  without the new owner's knowledge.
The victim is then either the new DNS name owner,
  who does not get customers,
  or customers that unwittingly use the wrong linked name.

\subsubsection{Linked Resource Squatting}\:
  occurs when the linked resource is not independent from the linked name,
  so a squatting attack extends to the linked resource.
(It looks the same as linked name squatting, \autoref{fig:linked_name_squatting_picture}.)
The adversary who first created the DNS name and linkage
  retains access to the linked resource
  even after losing control of the DNS name.

For example, Maven Central derives namespace access
  from domain ownership
  but does not revoke it when ownership ends.
The original owner can continue to publish packages
  under the namespace after the domain expires.
A subsequent owner who claims the same namespace
  inherits these packages
  and cannot remove them due to Maven's immutability guarantee~\cite{sonatype_immutability_nodate},
  leaving the namespace polluted
  with packages from a previous ownership epoch.

\subsubsection{Linked Name Takeover}\:
  attacks occur when an adversary acquires a DNS name
  after its previous owner loses control (\autoref{fig:linked_name_takeover_picture}),
  such as by \emph{dropcatching} it
  at the moment it is deleted
  from the registry after expiration~\cite{miramirkhani_panning_2018}.
The adversary then creates a new linkage
  pointing to a linked name under their control.
The adversary captures value that has accumulated
  around the DNS name:
  existing web links, bookmarks, payment references,
  or social media recognition.
The adversary can then impersonate the previous owner:
  victims who follow links using the DNS name,
  expecting to reach the original owner's
  Ethereum payment address,
  now reach the adversary instead.

The DNS ecosystem provides safeguards
  such as renewal grace periods and expiration notices,
  but these only help owners who intend to keep their domains.
Once a domain lapses,
  dropcatching is common~\cite{miramirkhani_panning_2018}.

The victim in a linked name takeover is the original creator of
  the DNS and linked names, and potentially new users attempting
  to access that name.

\subsubsection{Linked Resource Takeover}\:
  attacks occur when the linked resource
  is not independent from the linked name,
  so controlling the linked name suffices
  to control the linked resource
  and everything accumulated under it (\autoref{fig:linked_resource_takeover_picture}).

The consequences of a Linked Resource Takeover can be severe.
For example, Maven Central derives namespace access
  entirely from domain ownership:
  whoever proves control of \texttt{example.com}
  gains publishing rights to \texttt{com.example}.
An adversary who registers an expired domain
  gains publishing rights to a namespace
  that may contain years of trusted packages
  with many downstream dependents
  (the dark blue portion of linked name in \autoref{fig:linked_resource_takeover_picture}).
The adversary can then publish malicious new versions
  of those packages, indistinguishable from legitimate updates.
An analogous attack recently succeeded in the WordPress ecosystem,
  where an adversary purchased a widely used plugin company
  with over 20k active installations,
  and published new versions
  that included a backdoor~\cite{ginder_someone_2026,fadilpasic_wordpress_2026}.
Linked Resource Takeover enables the same outcome
  without requiring the original owner's cooperation.
This attack is particularly dangerous with
  modern build systems that automatically download
  the latest version of packages from the Internet.
Victims are any applications that build with compromised libraries,
  subject to remote code execution or exfiltration of secrets.

\subsection{Automatic Defenses Against Attacks}

Two factors help defend against some types of attacks:
  \emph{automatic zombie revocation}
and  \emph{linked resource independence}.

Some systems \emph{automatically} revoke linkages.
A simple method is to \emph{time out} the linked name,
  requiring the linkage to be regularly renewed.
TLS Certificates carry validity periods,
  and those have moved from years to months or even days~\cite{server_cert_wg_latest_2026,mcpherrin_6-day_2026}.
Timeouts provide a fail-safe window to safety
  (assuming users do not override timeout validation),
  and limit the impact of Bulk Creation.
However, they still leave the linked name vulnerable for some time.
Unfortunately, of the systems we consider, only Web PKI has timeouts.

Other systems may periodically confirm and invalidate zombie linkages.
However, if confirmation is done periodically, it is not immediate,
  and the delay between the end of a DNS ownership epoch
  and actual revocation creates a window
  during which these attacks remain possible.
None of the systems we consider periodically validates linkages.

\emph{Independence of Linked Resources} from the linkage
  prevents Linked Resource Squatting and Linked Resource Takeover attacks,
  since any new linkage will create a new linked resource.
Independent resources cannot directly accessed by the adversary,
  however, new users using the DNS and its linkage to find the
  linked name and resource may find the adversary's new linked resource
  instead of the victim's old resource.
If the user accepts this resource as valid, the could inadvertently
  interact with the adversary, perhaps installing their malware
  or sending funds to their crypto wallet.

The goal of our study is to understand how these aspects affect different attacks
  across different ecosystems.
%
%
%
%
%
%

 % \iffalse
 % \ifhasbluesky
 % \fi
 % \fi

%
\section{Methodology}
  \label{sec:methodology}

Our methodology follows four steps
  to evaluate how well an ecosystem integrates with DNS:
  (1)~identify all linkages in the ecosystem,
  (2)~determine the birth and death of each,
  (3)~map each linkage's birth to a DNS ownership epoch,
  and (4)~determine when that epoch began and ended.

Because inferring DNS ownership epochs (steps 3 and 4)
  is common to all integrations,
  we describe our approach for these steps first,
  then detail each integration's methodology individually.

\subsection{Inferring DNS Ownership Epochs}
  \label{sec:methodology/dns}

Determining whether a linkage is a zombie
  requires identifying the DNS ownership epoch
  during which the linkage was created,
  and whether that epoch has since ended.
The ideal source of this information would be
  the full registration history of the DNS name,
  but registries do not make it publicly available.

\subsubsection{Data Sources}.
We instead combine three complementary sources:
  current registration status from RDAP,
  historical TLD zone delegation data,
  and active DNS resolution scans.
Each varies in authority and coverage,
  so combining them requires care.

Our approach builds on three principles.

First, registration is a prerequisite for delegation,
  which in turn is a prerequisite for resolution.
Observing a domain as delegated or resolvable on a given day
  confirms it was registered on that day (ignoring transient states),
  though not which ownership epoch it belongs to.

Second, RDAP provides authoritative
  ownership epoch boundaries.
A positive RDAP response returns
  the start of the current ownership epoch,
  though not whether prior epochs existed.
A negative RDAP response confirms
  that no registration is currently active.

Third, sustained gaps in delegation data
  typically indicate domain expiry and deletion.
For gTLDs, a domain typically passes through
  auto-renew grace, redemption, and pending delete
  before its delegation is removed from the parent zone,
  a process taking about 80~days~\cite{affinito_domain_2022}.
A gap of 80~days or more in delegation
  is therefore strong evidence
  that the domain was fully deleted,
  though not authoritative:
  ccTLDs follow different lifecycles,
  and delegation lapses can occur
  without a change in registration.
When delegation data is unavailable,
  we fall back on gaps in active scan observations
  using the same threshold,
  though this is a weaker signal.

\subsubsection{Algorithm}.
Our algorithm combines these three principles
  to infer DNS ownership epochs.
We describe the algorithm briefly here;
  the full algorithm is in the appendix
  (\autoref{alg:appendix/epochs}) for space.

The algorithm infers registration intervals
  for a given domain from three inputs:
  daily TLD zone delegation observations ($Z$),
  daily active scan observations ($S$),
  and RDAP observations ($R$).
Two tunable parameters control sensitivity.
A delegation gap threshold ($t$)
  filters transient gaps in zone delegation,
  and a grace window ($g$)
  tolerates small misalignments
  between RDAP-reported and delegation-derived
  interval boundaries.

The algorithm operates in four phases.

In the first phase,
  the algorithm builds a unified bitset
  from daily zone delegation and active scan data,
  assigning $1$ to each day a domain was observed
  and $0$ otherwise.

In the second phase,
  candidate registration intervals are extracted
  as runs of consecutive $1$s in the bitset.
These intervals are initially \emph{open},
  meaning either end may be extended
  or merged with adjacent intervals
  in subsequent phases.

In the third phase,
  RDAP data refines the candidate intervals.
An authoritative registration date
  can \emph{close} the start of an interval,
  or split it if the date falls mid-interval,
  indicating a re-registration.
An authoritative negative response
  can prevent the merger of two adjacent intervals.

In the fourth phase,
  adjacent intervals are merged
  if the gap between them is less than $t$~days
  and no RDAP evidence prevents it.
We use $t=80$ days to align with the duration of
  the typical gTLD domain expiration process.

The intervals that remain after all four phases
  are output as inferred registration intervals.

Once the DNS ownership intervals are inferred,
  classifying a linkage as a zombie is straightforward:
  find the ownership interval active when the linkage was created
  and check whether it has since ended.
The same intervals yield additional metrics,
  such as zombie duration
  and domain age at the time of linkage creation.

\subsubsection{Specific Datasets}.
Our active scan data consists of daily NS and SOA queries
  for each DNS name corresponding to
  a linkage in the three ecosystems we study.

We source historical zone delegation data
  from dns.coffee~\cite{dnscoffee}.
This data spans 1291 TLDs from 2021 onward,
  though coverage varies by TLD.

Our RDAP data consists of 149M successful queries
  to 113M unique domains since 2023.
Most of these queries target newly registered domains observed in CT logs,
  but we query RDAP for every domain in our study at least once.

\subsubsection{Algorithm Limitations}.
Strictly speaking, our algorithm infers
  \emph{registration} epochs and not \emph{ownership} epochs,
  so zombies caused by within-registration
  ownership changes go undetected.
A domain that is sold or transferred without re-registration
  changes ownership without resetting its registration date,
  making such transitions invisible to our data sources.
Detecting registrant changes through RDAP
  is impractical because registrant fields
  are heavily redacted post-GDPR,
  and frequently obscured by privacy proxy services.

Our 80-day gap threshold is deliberately conservative
  and may under-count zombies,
  particularly domains that are dropcatched
  immediately after expiration.
For such cases we rely entirely on RDAP
  to detect the ownership change
  via a new registration date.
We prefer under-counting to over-counting,
  since classifying a live linkage as zombie
  is more harmful to our analysis
  than a missed zombie.

Finally, our algorithm operates at day granularity,
  so events within the same day are indistinguishable.
This is a deliberate choice
  matching the resolution of our data:
  zone files are published daily
  and our active scans run once per day.

These algorithmic limitations compound
  with the dataset limitations
  that we discuss next.

\subsubsection{Dataset Limitations}.
Each of our three data sources
  has limitations in coverage and accuracy.

RDAP, our most authoritative source,
  has the most uneven coverage.
Some TLDs lack RDAP servers entirely
  (such as \texttt{.ee}),
  others return responses that omit registration dates
  (such as \texttt{.de}),
  and rate limiting causes query failures
  even for TLDs that do support RDAP.

Zone delegation coverage is limited
  primarily to gTLDs participating in
  ICANN's CZDS program~\cite{iana_nodate},
  leaving most ccTLDs uncovered.
Even for covered TLDs,
  zone file update frequency varies,
  introducing timing uncertainty
  in when a delegation change becomes visible.
Zone files may also miss
  short-lived domains that are registered
  and removed between successive snapshots~\cite{sommese_darkdns_2024}.

Since our active scan coverage begins
  at the start of our data collection period,
  we have no visibility into earlier events.
Additionally, scan failures may reflect
  network-layer issues such as timeouts
  rather than changes in registration or delegation status.

We are in the process of obtaining ground-truth registration history
  from a major TLD
  to evaluate the accuracy of our inferred ownership epochs
  in light of these limitations.

\subsection{Measuring Linkages in Ecosystems}
  \label{sec:methodology/ecosystems}

Next, we describe how we identify linkages in each ecosystem,
  and how we determine their birth and death.

\textbf{Web PKI.}
The linkages we study in the Web PKI are domain-validated TLS certificates
  of \emph{newly registered} domains.
We focus on newly registered domains because
  they are more likely to be short-lived~\cite{sommese_darkdns_2024},
  and therefore more likely to produce zombies.

To collect certificates,
  we identify Fully Qualified Domain Names (FQDNs) under newly registered domains
  by monitoring Certificate Transparency (CT) logs
  via OpenINTEL's ZoneStream service~\cite{sommese_darkdns_2024},
  and contact each via TLS.
We exclude invalid certificates from our dataset
  by validating the presented certificate chain against Mozilla's Root CA bundle.

We define the birth of a certificate as its \texttt{notBefore} timestamp,
  and its death as the earlier of its \texttt{notAfter} timestamp
  or its revocation timestamp, if revoked.
We check for revocation first using CRLs to minimize traffic,
  and fall back to OCSP queries when CRLs are unavailable.

Over a period of six months (2025-10-01 to 2026-04-15),
  we collect 52.2 million certificates
  across 23.6 million newly registered domains.
This subset represents 2.7\% of 1.9 billion certificates
  issued during our study period~\cite{cloudflare_certificate_nodate}.

\textbf{ENS.}
Our ENS data collection differs
  between On-chain and Gasless linkages.

We identify On-chain linkages by inspecting Ethereum blockchain records.
Since the On-chain linkage process involves calls
  to ENS's \texttt{DNSRegistrar} smart contract,
  we collect all transactions to this contract
  using Etherscan~\cite{etherscan_etherscan_nodate},
  and decode them using the contract's ABI.
Successful linkages emit a \texttt{Claim} event,
  which we use to extract the list of linkages.
We gather 1,882 linkages since On-chain's inception in 2018.

We define the birth of an On-chain linkage
  as the timestamp of the block containing its \texttt{Claim} event,
  and its death as the timestamp of the \emph{next} \texttt{Claim} event (if any)
  for the same DNS name.
On-chain linkages do not expire and cannot be revoked;
  the only way to invalidate one is to overwrite it by repeating the linkage process.

Unlike On-chain linkages, no master list of Gasless linkages exists by design,
  so we discover them through active DNS scans.
We search OpenINTEL's forward DNS dataset~\cite{van_rijswijk-deij_high-performance_2016}
  for TXT records matching the format required by ENS Gasless~\cite{gregskrileth_gasless_2024}.
Since these scans cover only DNS names found in zone files and CT logs,
  our count (see \autoref{fig:ens_gasless_linkages}) is a lower bound.
Because Gasless linkages are validated using DNSSEC on each use,
  they cannot produce zombies by design.

 % \ifhasbluesky
 % \fi

\textbf{Maven Central.}
The linkages we study in the Maven Central are software namespaces.

We identify namespaces and their published package versions
  by crawling a Maven Central repository mirror~\cite{maven_central_central_nodate}.
We restrict our analysis to DNS-based namespaces using reverse-DNS convention
  (now mandatory),
  identifying 19.1M package versions across 31,853 namespaces.

Because namespace creation dates are not publicly available,
  we use the date of the first published package version
  as the birth of the namespace.
Due to Maven Central's immutability policy,
  namespaces do not expire,
  so we treat them as having no death.

\section{Characterizing Zombies}
  \label{sec:char-zombies}

We next characterize zombies using historic and new data.
We look at how often linkages become zombies
  (\autoref{sec:char-zombies/zombie-fraction}),
  and causes of creation (\autoref{sec:char-zombies/zombie-formation}).
We consider zombie longevity and revocation
  (\autoref{sec:char-zombies/zombie-duration}).
Due to space limitations, we report on time-to-linkage as a risk indicator
  in an appendix
  \autoref{sec:appendix/zombie-risk-reg-to-linkage}.

\subsection{What Fraction of Linkages are Zombies?}
  \label{sec:char-zombies/zombie-fraction}

\begin{figure}
\centering
\begin{subfigure}{0.46\textwidth}
  \includegraphics[width=\linewidth]{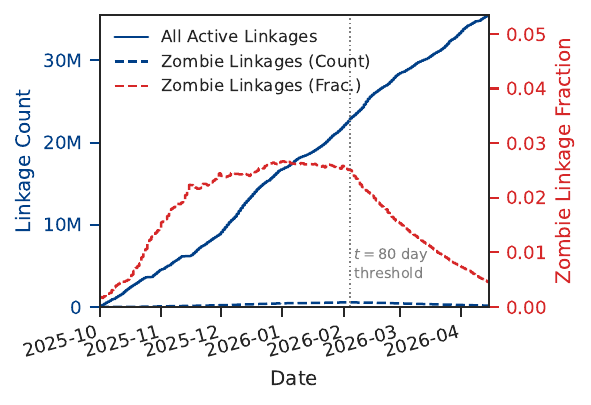}
  \caption{Web PKI}
  \label{fig:zombie_frac_webpki}
\end{subfigure}
\begin{subfigure}{0.46\textwidth}
  \includegraphics[width=\linewidth]{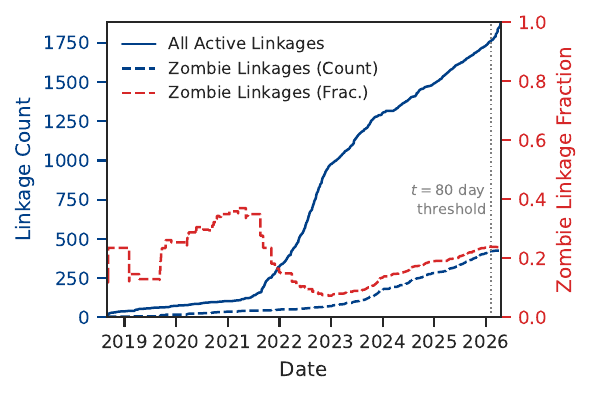}
  \caption{ENS (On-chain)}
  \label{fig:zombie_frac_ens}
\end{subfigure}
 % \ifhasbluesky
 % \fi
\begin{subfigure}{0.46\textwidth}
  \includegraphics[width=\linewidth]{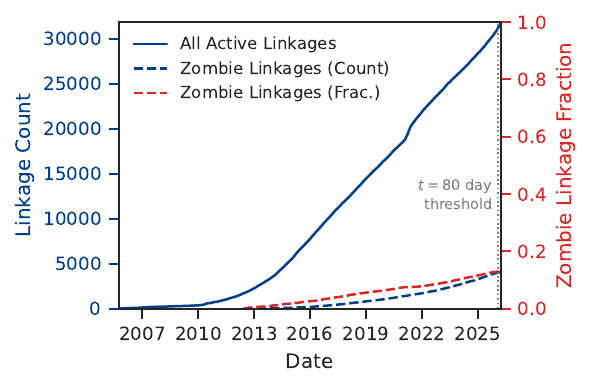}
  \caption{Maven Central}
  \label{fig:zombie_frac_maven}
\end{subfigure}
\caption{
  Zombie fraction over time across integrations.
  Each panel shows active linkage count (solid blue, left axis),
    zombie linkage count (dashed blue, left axis),
    and zombie linkage fraction (dashed red, right axis).
}
\label{fig:zombie_frac}
\end{figure}

\begin{figure}
  \begin{center}
    \includegraphics[width=0.8\linewidth]{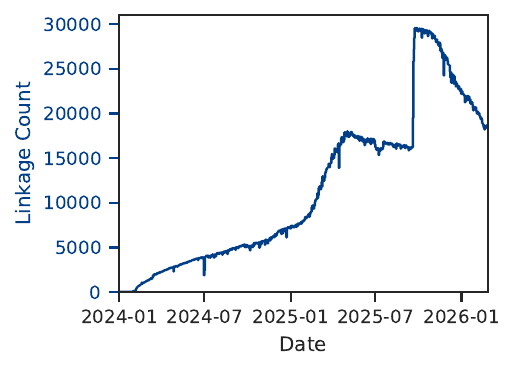}
  \end{center}
  \caption{
    Active ENS Gasless linkages over time.
  }
  \label{fig:ens_gasless_linkages}
\end{figure}

To establish the scale of zombies,
  we first identify the fraction of linkages
  that are active zombies at a given point in time.

\autoref{fig:zombie_frac} shows
  the fraction of linkages that are zombies (dashed red, right axis)
  along with counts of all (solid blue)
  and zombie (dashed blue) linkages
  in each ecosystem.
We also show our 80-day threshold
  (the gray dotted line, from \autoref{sec:methodology/dns});
  delegation and scan gaps after this point
  are too short to trigger zombie detection,
  though RDAP-based detection remains possible.

Most counts of active linkages grow over the measurement period,
  even though we track both births and deaths.
The reason for this growth varies by ecosystem.
For PKI, observed linkage births (52.2M total) outpace deaths (17.1M total);
  for ENS On-chain and Maven, linkages never die (see \autoref{sec:char-zombies/zombie-duration}).

\textbf{Web PKI.}
\autoref{fig:zombie_frac_webpki} shows
  zombie fraction and count among
  52.2M TLS certificates of newly created DNS names
  observed over six months (2025-10-01 to 2026-04-15),
  with a peak of 35.5M active certificates on a given day.

We see an increasing number of zombies over time,
  with a relatively small fraction of zombies (around 2.7\%).

We expect the fraction of zombie TLS certificates
  to remain steady over time,
  assuming the creation rate is stable and
  old zombies expire.
The apparent rise and fall of the zombie fraction
  is a measurement artifact,
  from newly registered DNS names having to die for zombies to form,
  and our 80-day classification threshold.
We discuss zombie formation
  and cleanup in
  \autoref{sec:char-zombies/zombie-formation} and \autoref{sec:char-zombies/zombie-duration}.

We see the zombie fraction is stable around 2.7\%
  from about 2025-11 to 2026-02, away from measurement transients.
Because newly created domains
  are disproportionately short-lived~\cite{sommese_darkdns_2024},
  this rate likely overstates the Web PKI-wide zombie fraction.
At the same time,
  since we track only a few domains
  (25M, while \texttt{.com} alone has over 150 million~\cite{dnscoffee}),
  the absolute number of zombies across the ecosystem
  is likely substantially larger.
Evaluating the full Web PKI population
  would require pairing CT log data
  with registration histories across all TLDs.

\textbf{ENS (On-chain).}
Since ENS On-chain linkages are valid indefinitely (until overwritten),
  we expect the number of zombies to grow.
\autoref{fig:zombie_frac_ens} confirms this hypothesis.
The number of zombies grow steadily over our 7 years of data.
The fraction of zombies varies from 7 to 30\%,
  dipping when there are many new registrations.
Of the 1,882 active ENS On-chain linkages
  as of April~2026,
  425 (23.8\%) are zombies.

Since 2023, the fraction of zombie fraction has been tending up
  as older linkages age.
Its nadir was 7\% in late 2022,
  with a surge in new ENS On-chain linkages.
Since ENS linkages have no expiration, we expect this growth to continue;
  in \autoref{sec:char-zombies/zombie-duration} we show that zombies have 
  never been cleared.

For comparison, \autoref{fig:ens_gasless_linkages} shows
  the growth of ENS Gasless linkages,
  which share the ENS ecosystem but are \emph{zombie-proof by design}
  because they are validated on each use.
Since late 2025, over 10k Gasless linkages have lost their backing TXT record
  and are no longer resolvable;
  had these been On-chain, every one would persist as a zombie indefinitely.

 % \ifhasbluesky
 % \fi

\textbf{Maven Central.}
Like ENS On-chain, Maven Central
  verifies namespace claims only once,
  so we expect Maven Central zombies
  to accumulate over time.

\autoref{fig:zombie_frac_maven} confirms this hypothesis:
  of 31,853 namespace claims as of March 2026,
  4,842 (15.2\%) are zombies,
  and both the absolute count and fraction seem to grow over time.

Maven's fraction of zombies is growing
  (15\% after 20~years),
  but slower than ENS On-chain
  (24\% after 7~years).
Both systems validate-once,
  but we suggest that software publishers
  place more value in continuity 
  than individuals experimenting with cryptocurrency.
However, Maven Central and ENS On-chain show
  that validate-once systems show much higher zombie fractions
  than systems with expiration (Web PKI). %

\subsection{What Causes DNS Names to Go Away?}
  \label{sec:char-zombies/zombie-formation}

\begin{figure}
  \centering
  \begin{subfigure}{0.42\textwidth}
    \includegraphics[width=0.95\linewidth]{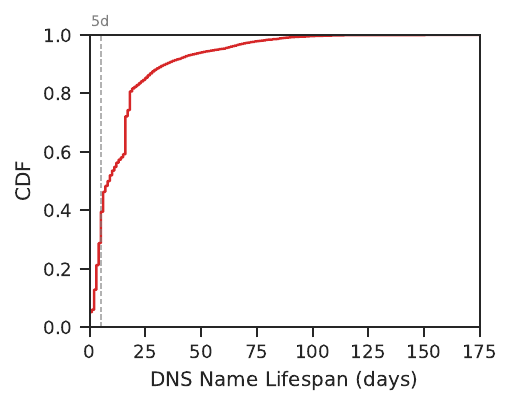}
    \caption{Web PKI}
    \label{fig:zombie_domain_lifespan_webpki}
  \end{subfigure}
 % \ifhasbluesky
 % \fi
  \begin{subfigure}{0.42\textwidth}
    \includegraphics[width=0.95\linewidth]{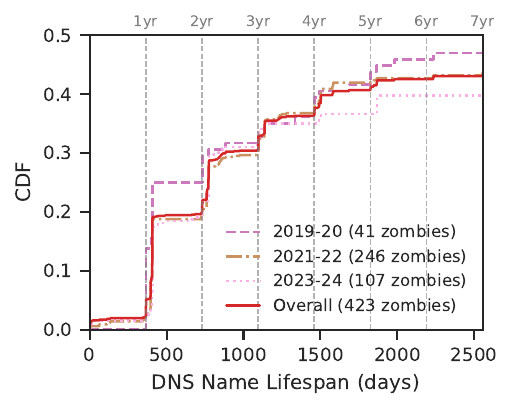}
    \caption{ENS (On-chain)}
    \label{fig:zombie_domain_lifespan_ens}
  \end{subfigure}
  \hfill
  \begin{subfigure}{0.42\textwidth}
    \includegraphics[width=0.95\linewidth]{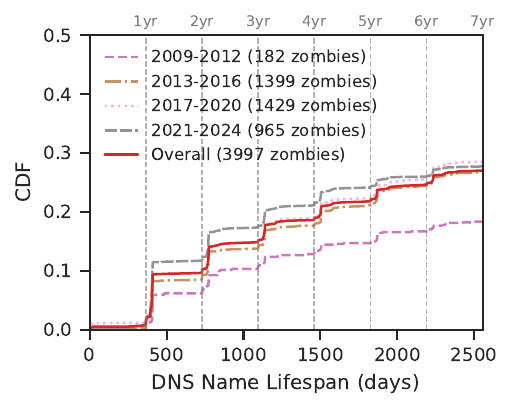}
    \caption{Maven Central}
    \label{fig:zombie_domain_lifespan_maven}
  \end{subfigure}
  \caption{
    Distribution of DNS name lifespans across integrations.
    Colored curves show per-cohort trends;
      the solid red curve shows the overall population.
    Vertical gray dotted lines mark
      5 days (AGP) for Web PKI,
      and years since registration for ENS On-chain and Maven Central.
  }
  \label{fig:zombie_domain_lifespan}
\end{figure}

Zombies are created when the DNS name goes away,
  so what causes DNS names to go away?
DNS names go away for four reasons.
Rarely, they are taken away in response to abuse or trademark disputes.
For domain tasting~\cite{dynadot_what_2025},
  DNS names are allocated conditionally and then removed
  by the registrant during a
  short Add Grace Period (typically 5 days).
Users occasionally voluntarily close a DNS name.
Most often, DNS names automatically expire
  after their registration period ends and they are not renewed.

We next look at DNS lifetimes for two reasons:
  very short lifetimes suggest Bulk Linked Name Creation
  (\autoref{fig:bulk_linked_name_creation_picture}), one of our attacks.
Second, understanding why zombies are created helps us evaluate
  whether zombies are due to changes in user choices
  (people are creating more zombies over time),
  or due to natural DNS aging breaking the linkage.

\textbf{Web PKI.}
Since we have only six months of DNS creation in 
  our Web PKI observation,
  we cannot observe zombie formation due to annual DNS expiry.
We can, however, assess early removals.

\autoref{fig:zombie_domain_lifespan_webpki}
  shows the distribution of DNS name lifetimes
  for 675k Web PKI zombies.
The dashed gray line marks
  ICANN's five-day Add Grace Period (AGP).

We see two modes in this graph: 5 days and about 16 days,
  and a tail out to 100 days.
We discuss each case next.

About two-fifths (39.6\%) of zombies die within five days,
  inside the Add Grace Period (AGP).
These zombies suggest domain tasting and Bulk Linked Name Registration.
An alternative is that these names are
  removed involuntarily 
  by registrars or registries
  in response to malicious activity~\cite{affinito_domain_2022},
  so our data establishes 40\% as an upper bound on 
  Bulk Linked Name Registration.

The second mode at 16~days falls outside the AGP,
  suggesting at least 20\% are removed due to abuse.

The upper limits of observed DNS name lifespan
  follow from our methodology.
All zombie-producing %
  DNS names 
  die within six months %
  because a name must die to produce a zombie,
  and our observation period is six months.

Since nearly 4 in 5 zombie-producing names die within 18~days,
  their certificates remain valid for the bulk of their issued lifetime,
  leaving plenty of time to exploit zombie certificates.
We examine certificate revocation in
  \autoref{sec:char-zombies/zombie-duration}.

 % \ifhasbluesky
 % \fi

\textbf{ENS (On-chain) and Maven Central.}
Since our ENS On-chain and Maven Central data
  spans years from each ecosystem's inception,
  we examine them together.

We show
  DNS name lifespans for linkages
  in ENS On-chain (\autoref{fig:zombie_domain_lifespan_ens}) and Maven Central (\autoref{fig:zombie_domain_lifespan_maven}),
  with each curve showing a cohort
  of linkages with about the same creation date
  (two-year groups for ENS On-chain, four-year groups for Maven Central).
The solid red line shows the overall trend
  across all cohorts.
To enable fair inter-cohort comparison,
  we include all linkages
  and use the Kaplan-Meier estimator~\cite{kaplan_nonparametric_1958}
  to account for right-censored observations.
The CDFs therefore do not reach 1.0.

In both ENS On-chain and Maven Central,
  zombie formation is clustered
  just past a multiple of years since DNS name registration.
The largest drop is always in the first year,
  with progressively smaller steps in subsequent years,
  consistent with a fixed annual non-renewal probability
  applied to a shrinking surviving population.
Domain renewal patterns show first-year renewal rates are the lowest,
  and renewal rates increase in subsequent years~\cite{sidn_why_2025}.
Regardless, this pattern strongly suggests
  that most zombies result from natural DNS expiration,
  rather than conscious DNS termination.

ENS On-chain shows little cohort effect,
  with per-year curves overlapping closely.
The curves start to diverge past year 5,
  but this spread reflects sparsity of data
  due to single-digit events post 5 years
  rather than a meaningful shift.

Maven Central, however, shows a clear cohort effect,
  with the earliest cohort  (2009--2012, the lowest dashed line)
  showing much
  longer-lived DNS names
  than others.
We suggest that
  early adopters of Maven Central were
  institutions or early Java users,
  both with long-held domains
  and important, long-lived software packages.
Later cohorts are more similar with each other,
  with less stable DNS names.

While DNS name expiry is the dominant driver
  of zombie formation in both ecosystems,
  we observe a few early DNS removals.
In ENS On-chain, 10 of 423 zombies (2.4\%)
  were formed by early removal,
  five of them within the AGP.
In Maven Central, 40 of 3,998 zombies (1.0\%)
  died within a year of DNS name registration,
  eight within the AGP.

These trends suggest two takeaways:
First, re-validation will be most efficient
  if it is done annually after DNS birth.
Second, zombies occur relative to DNS name age,
  so increasing fractions of zombies seen in \autoref{sec:char-zombies/zombie-fraction}
  reflect the age of each ecosystem
  and \emph{not} a change in user behavior.

\subsection{How Long do Zombie Linkages Persist?}
  \label{sec:char-zombies/zombie-duration}

\begin{figure}
  \begin{center}
    \includegraphics[width=0.75\linewidth]{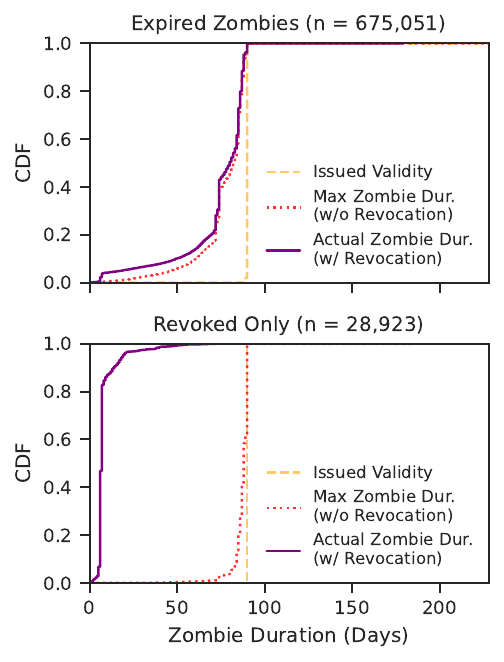}
  \end{center}
  \caption{
    Distribution of zombie duration
      for expired Web PKI certificates.
    The top panel shows all zombies;
      the bottom panel shows only those
      whose certificates were revoked.
  }
  \label{fig:webpki_zombie_duration}
\end{figure}

 % \ifhasbluesky
 % \fi

Given zombies exist in all studied ecosystems (\autoref{sec:char-zombies/zombie-fraction}),
  we now ask:
  how long do zombies persist after their linkage is broken?
The longer a zombie remains active,
  the longer an adversary has to carry out the attacks
  described in
  \autoref{sec:problem_statement/attacks}.

\textbf{Web PKI.}
The top panel of \autoref{fig:webpki_zombie_duration}
  shows the distribution of zombie duration
  for zombie TLS certificates that expired or were revoked
  during our observation period.
We exclude certificates still valid at the observation cutoff
  because they may yet be revoked before their expiry.

We find 675k certificates that meet this criteria,
  and the lifetime is 
  dominated by Let's Encrypt's 90-day certificates.
This mode is in part because
  our six-month-long observation 
  is too brief to capture both zombie birth and death
  for certificates with a 1-year expiration.
We are in the process of getting multi-year historic data
  that will see these zombies.

\autoref{fig:webpki_zombie_duration} shows
  the certificate validity period at issuance (dashed orange),
  which caps zombie lifetime and has
  almost all probability mass at 90 days.
The two curves show
  remaining validity after DNS name expiration (dotted red),
  the maximum zombie duration assuming no revocation,
and actual observed zombie duration (solid purple, left-most),
  which includes both expiration and revocation.

Most zombies persist for nearly
  the full validity period of 90~days:
  the median is 75~days,
  with 40\% lasting 80 days or more,
  and only 10\% under 45~days.
Our measurement precision is $\pm 24$\,h,
  which underestimates zombie lifetime,
  so this trend of ``near 90 days'' is very strong.

Shorter expiration times would shrink zombie duration.
Let's Encrypt plans to reduce
  default certificate expiration to 45~days
  by 2027~\cite{mcpherrin_decreasing_2025},
  and the CA/Browser Forum mandates
  a maximum validity of 47~days
  by 2029~\cite{server_cert_wg_latest_2026}.
Both changes would cut median zombie lifetime,
  and therefore the number of active zombies, roughly in half.

Revocation, provides little benefit
  because it is almost never used in practice,
  as shown by the similarity of curves with and without revocation (solid and dotted).

\emph{Web PKI Revocation:}
Although rarely used, is revocation important when it is used?
We consider the subset of zombie certificates that were revoked.

The bottom panel of \autoref{fig:webpki_zombie_duration}
  examines the lifetime of the 28.9k (4.3\% of all 675k) zombies that were revoked.
DNS name expiration rarely results in certificate revocation.

The revoked cohort show a large reduction in zombie lifetime:
  the median reduction is 82~days (92\%),
  eliminating nearly the entire zombie window.
Thus
  when revocations occur, they occur very soon after certificate creation.

Nearly all zombies (96\%), however, are never revoked,
  suggesting revocation today is not an effective defense against zombies.
While encouraging revocation might reduce vulnerability,
  this data confirms the benefits of shorter expiration times
  to reduce the window of zombie vulnerability.

\textbf{ENS (On-chain).}
ENS On-chain zombies persist until the new DNS owner
  actively reclaims the name with a new linkage.
Our data shows that \emph{none} of the 1,882 ENS On-chain zombies have been reclaimed,
  meaning \emph{every} zombie that has ever formed remains active today.
The median ENS On-chain zombie today is 1.9 years old. %

Short of the new DNS name owner repeating the linkage process,
  there is currently no mechanism to revoke an ENS On-chain zombie.
An NSEC3-based method for invalidating linkages
  via negative proofs once existed,
  but was removed in 2022~\cite{arachnid_commit_2022}
  without ever being used in practice.

This permanence of zombies makes ENS On-chain particularly susceptible
  to Linked Name Squatting.
An adversary's zombie linked name remains resolvable
  indefinitely at no ongoing cost,
  even if the DNS name is later re-registered by a new owner.
We discuss this attack further in \autoref{sec:attacks/overall}.

 % \ifhasbluesky
 % \fi

\textbf{Maven Central.}
Maven Central zombies are permanent.

Unlike ENS On-chain, where a new DNS name owner may overwrite a zombie
  by repeating the linkage process,
  Maven Central's immutability guarantee
  prevents any modification or removal of
  published package versions~\cite{sonatype_immutability_nodate}.
(In practice, however, Maven Central has removed packages
  containing malicious code in the past~\cite{sonatype_security_research_team_sonatype_2021}.)
A zombie namespace and all packages
  published under it therefore persist indefinitely,
  regardless of DNS ownership changes.

Our data shows that, as of April~2026,
  Maven Central has 4,842 zombie namespaces,
  with a median zombie age of 3.5~years.

Because Maven Central namespaces are immutable
  and publishing rights rest with the linked name owner,
  zombie permanence enables both Linked Resource Squatting and Takeover.
A former DNS name owner can continue publishing
  under a zombie namespace (Linked Resource Squatting)
  because Maven Central does not re-verify DNS ownership.
A new DNS name owner can take over the namespace
  and publish compromised versions (Linked Resource Takeover),
  though namespace takeovers go through manual review.
We discuss these attacks further in \autoref{sec:attacks/maven}.

\section{Identifying Attacks Involving Zombies}
  \label{sec:attacks}

\newcommand{\preventedS}{$\times$}
\newcommand{\noevidenceS}{$\circ$}
\newcommand{\insufficientS}{?}
\newcommand{\consistentS}{\checkmark}
\newcommand{\escalatesS}{$\uparrow$}

\newcommand{\prevented}{\cellcolor[HTML]{a0ffa0}\preventedS}
\newcommand{\noevidence}{\cellcolor[HTML]{a0ffff}\noevidenceS}
\newcommand{\insufficient}{\cellcolor[HTML]{ffff80}\insufficientS}
\newcommand{\consistent}{\cellcolor[HTML]{ff8080}\consistentS}
\newcommand{\escalates}{\cellcolor[HTML]{ffc0c0}\escalatesS}

\begin{table}
  \centering
  \resizebox{\columnwidth}{!}{%
  \begin{tabular}{l|ccc}
      & \textbf{Web}
      & \textbf{ENS}
      & \textbf{Maven} \\
      & \textbf{PKI}
      & \textbf{(On-ch.)}
      & \textbf{Central} \\
      \hline
      Bulk Linked Name Creation
      & \consistent{} \autoref{sec:attacks/webpki}
      & \noevidence{}
      & \noevidence{} \\
      Linked Name Squatting
      & \consistent{} \autoref{sec:attacks/webpki}
      & \consistent{}
      & \escalates{} \\
      Linked Resource Squatting
      & \prevented{}
      & \prevented{}
      & \consistent{} \autoref{sec:attacks/maven} \\
      Linked Name Takeover
      & \insufficient{}
      & \noevidence{}
      & \escalates{} \\
      Linked Resource Takeover
      & \prevented{}
      & \prevented{}
      & \consistent{} \autoref{sec:attacks/maven} \\
  \end{tabular}
  }
  \caption{
    Attack indicators across ecosystems.
    Design properties:
      \preventedS{} prevented by design,
      \escalatesS{} escalates to resource attack.
    Data findings:
      \noevidenceS{} no evidence of attack,
      \insufficientS{} insufficient data to determine,
      \consistentS{} actively available for exploitation.
  }

  \label{tab:attack_indicators}
\end{table}

In \autoref{sec:char-zombies} we show
  zombies \emph{exist}, but are they exploited?
We next use data for each ecosystem
  to evaluate if the attacks described in \autoref{sec:problem_statement/attacks}
  are \emph{actively available for exploitation}.
Although fully confirming exploitation requires
  data we do not have,
  we can show indicators consistent with an attack, and
  for Maven Central  (\autoref{sec:attacks/maven}),
  resource modification.

\subsection{Attack Evaluation}
	\label{sec:attacks/overall}

\autoref{tab:attack_indicators} summarizes
  our findings so far.

In the best case, \emph{attacks are prevented by  design} in some ecosystems
  (marked with \preventedS~and green) in \autoref{tab:attack_indicators}.
Resource attacks
  (Linked Resource Squatting and Linked Resource Takeover)
  are prevented in Web PKI and ENS On-chain %
  because their linked resources 
  require a separate cryptographic key,
  independent of the linked name (see \autoref{tab:terms}).

Our data shows \emph{no evidence of several attacks
  that are otherwise viable} by design
  (labeled~\noevidenceS~and cyan).
We find no evidence of Bulk Linked Name Creation in ENS On-chain or Maven Central,
  where very few zombies form within the Add Grace Period
  (5 of 423 for ENS On-chain, 8 of 3,998 for Maven Central;
  see \autoref{fig:zombie_domain_lifespan}).
 % \ifhasbluesky
 % \fi
No ENS On-chain zombie has ever been overwritten
  by a new DNS name owner
  (\autoref{sec:char-zombies/zombie-duration}),
  indicating that Linked Name Takeover has not occurred.

\emph{One attack--ecosystem pair
  lacks sufficient data to draw a conclusion} (marked~?).
Our Web PKI dataset contains only newly registered DNS names,
  so we cannot observe Linked Name Takeovers
  which require re-registration of an existing name.
 % \ifhasbluesky
 % \else
We will review this attack when we have more data.
 % \fi

\emph{The remaining attacks are both viable by design
  and actively available for exploitation}
  based on our data (marked~\consistentS{}).
In Web PKI, nearly 300k newly created DNS names
  with TLS certificates die within the Add Grace Period
  (\autoref{sec:char-zombies/zombie-formation}),
  leaving zombie certificates available
  for Bulk Linked Name Creation.
Linked Name Squatting in Web PKI
  is supported by 580k active zombie certificates
  that persist for most of their validity period
  (\autoref{sec:char-zombies/zombie-duration}).
In ENS On-chain, 456 zombies persist indefinitely with none ever cleared
  (\autoref{sec:char-zombies/zombie-duration}),
  each available for Linked Name Squatting.
In Maven Central, 4,842 zombie namespace claims
  persist permanently due to immutability guarantees,
  leaving them available for both
  Linked Resource Squatting and Linked Resource Takeover
  given Maven Central's lack of resource independence.

Although our data
  shows evidence of zombie that could be exploited,
  that presence consistent with an attack
  does not imply \emph{actual} exploitation.
We now look for stronger evidence
  of possible attacks in Web PKI and Maven Central.
We also examined ENS On-chain for evidence of exploitation,
  but the blockchain does not contain enough data for 
  us to confirm.
Linked Name Squatting requires that funds are sent to the zombie linked name,
  but Ethereum transactions do not record whether
  funds were sent via the linked name (potentially misdirected)
  or via the raw wallet address (legitimate).

\subsection{Attacks on Web PKI}
  \label{sec:attacks/webpki}

\begin{figure}
  \centering
  \includegraphics[width=0.7\linewidth]{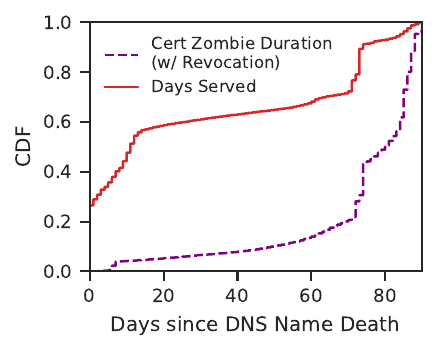}
  \caption{
    Duration Web PKI zombie certificates are served
      after DNS death (solid red line).
  }
  \label{fig:webpki_zombie_cert_usage}
\end{figure}

In \autoref{sec:char-zombies/zombie-fraction} we show
  that there are many zombies in the PKI ecosystem.
We next go two steps further in the attack chain
   to show that these zombies
  are active on servers,
  and that zombie certificates sometimes overlap with new domain owners.
Evaluation of server contents for malware is future work.

\subsubsection{Zombie PKI Certificates are Served}:
First we show that zombie certificates are actually served,
  posing a viable threat.

To determine how long zombie certificates are served, %
  we contact each TLS server daily at its last known IP address,
  checking if it presents the now-zombie certificate.
We study servers until the certificate expires or is revoked,
  since validating clients will then reject the certificate.
Our measurements provide a lower-bound on this vulnerability
  since we contact only the last known IP,
  missing cases where the certificate is served at a new address.

\autoref{fig:webpki_zombie_cert_usage} shows the results
  for 675k zombie certificates during our six-month observation period.
We compare how long certificates
  are actively served by a TLS server after DNS name death (the top, solid red line).
We also show zombie duration,
  from DNS name death to certificate expiry or revocation (the lower dashed purple line),
  as in \autoref{fig:webpki_zombie_duration},
  an upper-bound on days served.

While a quarter of zombie certificates are never served once they are zombies,
  about one-third (32\%, or 215k)  are served for 60~days or more.
An adversary who can direct victims to such a server
  can impersonate the domain with a certificate
  that browsers accept as valid.

\subsubsection{Served Zombie PKI Certificates Overlap with New DNS Owners}:
Next we show that some zombie certificates that are served
  overlap with new DNS owners.
This overlap is a potential example of malicious Linked Name Squatting 
  in the Web PKI ecosystem.

Of 675k zombie certificates,
  the validity period of 33k zombies (4.9\% of all zombies)
  overlaps with a new DNS name registration,
  and 7.3k of these (1.1\% of all zombies)
  are served past re-registration.
Both figures account for revocation.

\autoref{fig:appendix/webpki_zombie_served_past_rereg} (in appendix) shows
  how long these 7.3k zombie certificates
  continue to be served after their DNS names are re-registered:
  median certificate is served for 49~days past re-registration,
  a window during which an adversary
  can impersonate the new DNS name owner.

Revocation is one defense against zombie certificate misuse.
We therefore compare revocation rates for zombie certificates
  whose DNS names are re-registered (33k)
  with those that are not (642k).
We see \emph{much} higher revocation rates
  for zombies whose DNS names are re-registered (11.6\%)
  than for those that are not (3.9\%).
This $3\times$ higher fraction of revocations
  suggests that concern about zombie certificates results in action,
  yet the fact that 8 in 9 are never revoked
  and only go away on expiration suggests that considerable risk remains.

\subsection{Attacks on Maven Central}
  \label{sec:attacks/maven}

\begin{table}
\centering
\renewcommand{\arraystretch}{1.1}
 \resizebox{\columnwidth}{!}{
  \begin{tabular}{l r r r}
      \textbf{Category} & \textbf{Count} & \multicolumn{2}{c}{\textbf{(\%)}} \\
      Maven namespaces & 31\,853 & 100.0 \\
      \quad Live namespaces & 27\,011 & 84.8 \\
      \quad Zombie namesp. (unk. zombie start) & 789 & 2.5 \\
      \hline
      \quad Zombie namesp. (known zombie start) & 4\,053 & 12.7 & \emph{100.0} \\
      \qquad      No changes while zombie & 3\,506 & 11.0 & \emph{86.5} \\
      \qquad      New versions published while zombie & 547 & 1.7 &\emph{13.5} \\
      \quad\qquad DNS Name not re-registered & 257 & 0.8 &\emph{6.3} \\
      \quad\qquad DNS Name re-registered & 290 & 0.9 & \emph{7.2} \\
      \qquad\qquad No changes after re-registration & 76 & 0.2 &\emph{1.9} \\
      \qquad\qquad New versions after re-registration & 214 & 0.7 & \emph{5.3} \\
  \end{tabular}
 }
\caption{
  Zombies in Maven Central namespace.
}
\label{tab:maven_zombie_activity}
\end{table}

To study Linked Resource Squatting and Takeover in Maven Central,
  we examine whether new package versions are published
  after a namespace becomes a zombie in \autoref{tab:maven_zombie_activity}.
This table breaks down 31,853 Maven Central namespaces
  based on their zombie status and latest publishing activity.

Of 4,053 zombie namespaces with known zombie start dates,
  \emph{547 (13.5\%) publish new package versions after zombie birth}.
Any application that depends on these packages
  may unknowingly pull code published
  by someone who no longer controls the DNS name.

Of those 547 namespaces, 290 have since had their DNS names re-registered,
  of which 214 continue publishing new versions after re-registration.
These 214 namespaces pose the most risk:
  applications that pull new versions from them
  may unknowingly incorporate compromised code,
  exposing their users to attacks.
Recall that our algorithm detects re-registrations
  via RDAP registration date changes
  or delegation gaps of 80~days or more,
  so same-owner registrar transfers
  are not counted (\autoref{sec:methodology/dns}).

Confirming whether these zombie namespaces
  are active attack vectors will require
  comparing package signing keys and contents across versions.
Nevertheless, our findings demonstrate
  that resource attacks on Maven Central are feasible.
Following MavenGate~\cite{oversecured_introducing_2024},
  Sonatype stated in 2024 that accounts associated with expired domains
  were disabled~\cite{fox_sonatypes_2024},
  but our data shows zombie namespaces
  with publishing activity as recent as 2026,
  indicating this defense may be inadequate.
It also remains to be seen whether
  manual validation as a defense against
  re-claiming existing namespaces~\cite{fox_sonatypes_2024}
  can withstand well-resourced adversaries.

\section{Recommendations for DNS Integrations}
  \label{sec:discussion}

Our findings suggest several design principles
  for current and future DNS integrations.

The most effective defense against zombies
  is to \emph{validate linkage status on any use} of a linked name.
ENS Gasless demonstrates this approach:
  it validates linkages on-demand using DNSSEC,
  bounding the zombie window to the DNS cache duration
  (typically minutes).

Limiting linkage lifetime with a time-to-live
  is a great second defense,
  because it guarantees a limit on the duration which a name can be exploited.
The Web PKI ecosystem does this well,
  with every certificate including a time-to-live.
Web PKI has experimented with many techniques
  (revocation lists~\cite{Cooper08a},
  OCSP~\cite{Santesson13a} are two),
  but shortening certificate validity has been very effective, 
  prompting shortening certificate lifetime from years
  to a proposed 47 days~\cite{server_cert_wg_latest_2026},
  with Let's Encrypt recently trying 6 days~\cite{mcpherrin_6-day_2026}.

We recommend Resource Independence,
  where the linked resource is protected by an ecosystem-specific mechanism.
Resource Independence provides defense in depth,
  separating control of the linked resource with additional mechanisms
  (private key for PKI and ENS, for example).
Of the three integrations we study,
  only Maven Central is vulnerable to resource tampering
  (injection of new versions)
  through linked name control alone.

Finally, 
  our findings provide heuristics for prioritizing active re-validation
  to detect and mitigate zombies,
  for ecosystems that do not have other mitigations.
Zombie formation concentrates
  around the first few registration anniversaries
  (\autoref{sec:char-zombies/zombie-formation}),
  and linkages made soon after DNS registration
  are more likely to become zombies
  (\autoref{sec:appendix/zombie-risk-reg-to-linkage}).

 % \iffalse
 % \fi

%
\section{Conclusion}
  \label{sec:conclusion}
We study synchronization of DNS integration across three ecosystems
  and find that zombie linkages exist in all of them,
  but at sharply different rates
  driven by ecosystem design choices.
Our analysis shows that zombie linkages
  are actively available for exploitation
  in Web PKI and Maven Central.
Based on these findings,
  we recommend that new integrations
  treat DNS ownership change as a first-class event
  by validating linkages on every use
  and bounding linkage lifetimes.
As more ecosystems anchor trust in DNS names,
  the synchronization problem we characterize here
  will only grow in scope and consequence.

\begin{acks}

We sincerely thank Ian Foster of dns.coffee
  for providing access to historical TLD zone delegation data.
We thank Wes Hardaker (Google)
  for starting this work when he was at USC/ISI,
  and for his technical comments on its evolution.
We thank 
  Swapneel Sheth and Andrew Kaizer (Verisign),
  and
  Rick Wilhelm and Suzanne Woolf (PIR),
  for their technical comments 
  on early versions of this research.
This research was partially supported by gifts from
  Verisign and Public Interest Registry (PIR).
John Heidemann's work was support in part by NSF awards 
  CNS-2319409, %
  OAC-2530698, %
  and 2453092. %

\end{acks}

\label{page:last_body}

\printbibliography

\appendix

\section{Ethics}
    \label{sec:appendix/ethics}

In evaluating the ethics of our work,
  we considered data sources and personal data,
  and vulnerability assessment and the potential need for disclosure.
Overall, we do not believe our work poses any ethical challenges
  given the steps we have taken.

This paper used a range of non-personal data
  including DNS zone files, RDAP queries, TLS certificates,
  Maven Central repository metadata~\cite{maven_central_central_nodate},
  and Ethereum blockchain records.
Much of this data is available publicly.
Some data was aggregated by third-parties
  (dns.coffee~\cite{dnscoffee} and OpenINTEL~\cite{van_rijswijk-deij_high-performance_2016,sommese_darkdns_2024})
  and used with their permission.
We will provide data that we collected for this paper
  available at no cost upon publication,
  and provide pointers to third-party aggregations that we used.

Our work uses no personally identifiable information that is not already public
  (for example, in the Ethereum blockchain).
We therefore do not consider it to pose any privacy risks to individuals.

When querying public web endpoints
  (such as RDAP servers and Maven Central),
  we rate-limited our requests,
  respected throttling signals,
  and cached results wherever possible to avoid redundant queries.

Our work assesses security threats
  and identifies new classes of attacks on DNS integrations.
We discuss the security implications of the design decisions of
  different ecosystems.
However, we do not identify specific vulnerabilities
  that might require software patches or coordinated disclosure.

\section{Algorithm to Infer DNS Ownership Epochs}
  \label{section:appendix/alg-detailed}

\begin{algorithm*}
  \LinesNotNumbered

  \caption{Inferring DNS name ownership epochs.}
  \label{alg:appendix/epochs}

  \KwIn{%
    \begin{tabular}[t]{@{}l@{}}
      TLD zone delegation observations $Z$ \\
      Active scan observations $S$ \\
      RDAP observations $R$ \\
      Delegation gap threshold $t$ (default: 80 days)\\
      Grace window $g$ (default: 2 days)
    \end{tabular}%
  }
  \KwOut{Inferred registration intervals $\mathcal{I}$}
  \BlankLine

  \tcc{Phase 1: Bitset of delegation \& scan}
  $B_z \gets \textsc{DayObsBitset}(Z)$; \ \ $B_s \gets \textsc{DayObsBitset}(S)$\;
  $B \gets B_z \lor B_s$\;
  \BlankLine

  \tcc{Phase 2: Extract intervals}
  $\mathcal{I} \gets \textsc{ExtractRuns}(B)$\tcp*{all ends open}
  \BlankLine

  \tcc{Phase 3: RDAP refinement}
  $R^{+} \gets$ positive observations in $R$, keeping latest $o.\mathit{time}$ per $o.\mathit{reg}$\;
  $R^{-} \gets$ negative observations in $R$\;
  \ForEach{$o \in R^{+}$ in order of $o.\mathit{reg}$}{
    $I_{\mathit{reg}} \gets$ interval containing $o.\mathit{reg}$, else create new\;
    $I_{\mathit{obs}} \gets$ interval containing $o.\mathit{time}$, else create new\;
    \uIf{$o.\mathit{reg} - I_{\mathit{reg}}.\mathit{start} \leq g$}{
      make $I_{\mathit{reg}}.\mathit{start}$ closed\;
    }
    \Else{
      split $I_{\mathit{reg}}$ at $o.\mathit{reg}$ with new closed start\;
    }
    \ForEach{$I \in [I_{\mathit{reg}}, \mathrm{prev}(I_{\mathit{obs}})]$}{
      $I.\mathit{mergeNext} \gets \text{True}$\;
    }
  }
  \ForEach{$o \in R^{-}$ in order of $o.\mathit{time}$}{
    \If{$o.\mathit{time}$ is not within $g$ days of any interval}{
      make start of next interval closed (if any)\;
    }
  }
  \BlankLine

  \tcc{Phase 4: Merge}
  \ForEach{adjacent $(I_i, I_{i+1}) \in \mathcal{I}$}{
    \If{$I_{i+1}.\mathit{start}$ is not closed \textbf{and} ($I_{i+1}.\mathit{start} - I_i.\mathit{end} < t$ \textbf{or} $I_i.\mathit{mergeNext}$)}{
      merge $I_i$ and $I_{i+1}$\;
    }
  }
\end{algorithm*}

\autoref{alg:appendix/epochs} shows the algorithm we use
  to infer DNS ownership epochs.
We discuss this algorithm in \autoref{sec:methodology/dns}.

\section{Does Time from Registration to Linkage Predict Zombie Risk?}
  \label{sec:appendix/zombie-risk-reg-to-linkage}

\begin{figure*}
\centering
\begin{subfigure}{0.48\textwidth}
  \includegraphics[width=\linewidth]{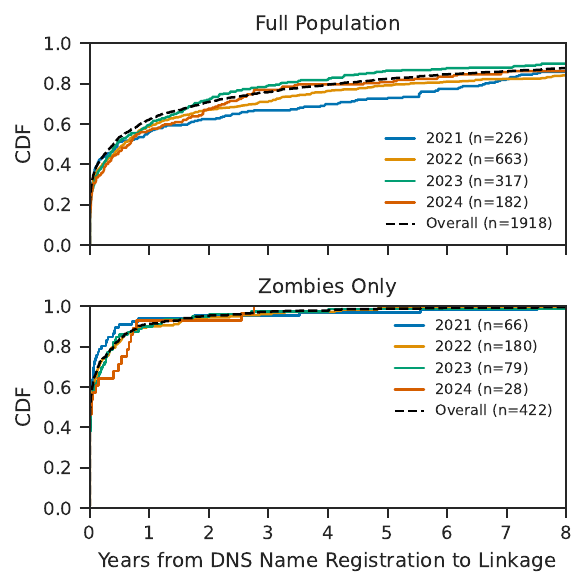}
  \caption{ENS (On-chain)}
  \label{fig:reg_to_linkage_ens}
\end{subfigure}
\hfill
\begin{subfigure}{0.48\textwidth}
  \includegraphics[width=\linewidth]{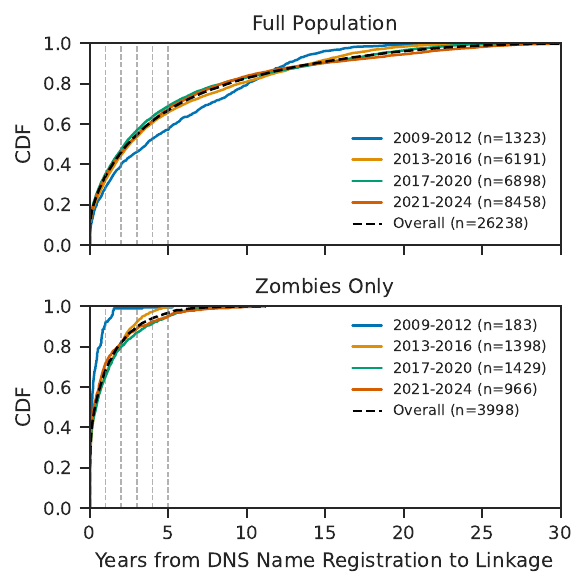}
  \caption{Maven Central}
  \label{fig:reg_to_linkage_maven}
\end{subfigure}
\caption{
  Time from DNS name registration to linkage creation
    for ENS On-chain (left) and Maven Central (right).
  The top panels show the full population;
    the bottom panels show only zombies.
  Colored curves show per-cohort trends;
    the dashed black curve shows the overall population.
  Vertical gray dotted lines mark yearly boundaries.
}
\label{fig:appendix/reg_to_linkage}
\end{figure*}

We study whether linkages that eventually become zombies
  are created sooner after DNS name registration
  than the general population.
Users who link an established DNS name
  may do so at any point in its lifetime,
  but users who register a name specifically to create a linkage
  would do so soon after registration.
If the two populations differ,
  registration-to-linkage timing may be used
  as a practical signal for identifying
  zombie-prone linkages early.

We limit this study to ENS On-chain and Maven Central.
Our Web PKI dataset contains only newly created domains,
  so the registration-to-linkage gap is short by construction.

\autoref{fig:appendix/reg_to_linkage}
  plots the years from DNS name registration to linkage creation
  for ENS On-chain (left) and Maven Central (right),
  for the full population (top) and zombies only (bottom).
The figure visually supports our hypothesis:
  the fraction of linkages created immediately
  after registration is nearly double for zombies
  compared to the general population
  (55\% vs.\ 30\% for ENS On-chain,
  30\% vs.\ 15\% for Maven Central).

To confirm this difference statistically,
  we apply a Mann-Whitney U test~\cite{mann_test_1947}
  comparing registration-to-linkage duration
  between zombie and non-zombie linkages.
A significant result ($p < 0.05$) means that zombie linkages
  are systematically created sooner after registration
  than non-zombie linkages.

Both ecosystems confirm the hypothesis
  ($p < 10^{-63}$ for ENS On-chain, $p \approx 0$ for Maven Central).
Zombie linkages are created far sooner after registration,
  with median gaps of 4~days (zombie) compared to
  292~days (non-zombie) for ENS On-chain
  and 123~days (zombie) compared to 1,169~days (non-zombie) for Maven Central.

The gap between DNS name registration
  and linkage creation is therefore
  a practical predictor of zombie risk,
  and ecosystems could use it as a heuristic
  to periodically re-validate zombie-prone linkages.

\section{Days Web PKI Zombies are Served Past DNS Re-registration}
  \label{sec:appendix/webpki_zombie_served_past_rereg}

\begin{figure*}
  \centering
  \includegraphics[width=0.35\linewidth]{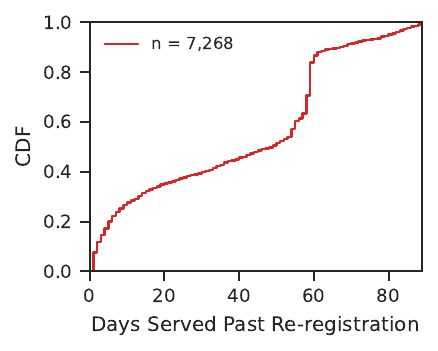}
  \caption{
    Duration Web PKI zombie certificates are served
      after their DNS names are re-registered.
  }
  \label{fig:appendix/webpki_zombie_served_past_rereg}
\end{figure*}

\autoref{fig:appendix/webpki_zombie_served_past_rereg} shows
  the distribution of duration 7.3k Web PKI zombie certificates are served
  after their DNS names are re-registered.
We discuss this figure in \autoref{sec:attacks/webpki}.

\label{page:last_page}

\end{document}